\documentclass[12pt]{iopart}
\usepackage{units}
\usepackage{graphicx}
\usepackage{booktabs}
\usepackage{subeqnarray}
\usepackage{cite}
\usepackage{epstopdf}
\usepackage{citesort}
\usepackage[american]{babel}
\usepackage{graphicx}
\usepackage{braket}
\usepackage{bm}
\usepackage{amssymb}

\def \grad {{\rm grad}}
\def \mod {\,{\rm mod}\, }

\def \be {\begin{equation}}
\def \ee {\end{equation}}
\def \ba {\begin{eqnarray}}
\def \ea {\end{eqnarray}}

\begin{document}

\title[Flat-face invisibility cloaks]{Flat-face approximations of invisibility cloaks with planar metamaterial layers}

\author{Oliver Paul$^{1}$, Yaroslav Urzhumov$^{2}$, Christoffer Elsen$^{1}$, David Smith$^{2}$, and Marco Rahm$^{1,3}$ }

\address{\vspace{5 pt}$^1$Department of Physics and Research Center OPTIMAS, University of Kaiserslautern, 67663 Kaiserslautern, Germany\vspace{5 pt}\\
 $^2$Center for Metamaterials and Integrated Plasmonics, Pratt School of Engineering,
Duke University, Durham, NC 27708, USA\vspace{5 pt}\\
$^3$Fraunhofer Institute for Physical Measurement Techniques IPM, 79110 Freiburg, Germany }
\ead{paul@physik.uni-kl.de}

\begin{abstract}
Transformation optics (TO) is a powerful tool for the design of artificial materials with unprecedented optical properties. General TO media are demanding, requiring spatially varying constitutive tensors with both anisotropic electric and magnetic response. Though metamaterials have been proposed as a path to achieving such complex media, the required properties corresponding to the most general transformations remain elusive even in metamaterials leveraging state-of-the-art fabrication methods.
Fortunately, in many situations the most significant benefits of a TO medium can be obtained even if approximations to the ideal structures are employed. Here, we propose the approximation of TO structures of arbitrary shape by faceting, in which curved surfaces are approximated by flat metamaterial layers that can be implemented by standard fabrication and stacking techniques. We illustrate the approximation approach for the specific example of a cylindrical ``invisibility cloak". First, we introduce a numerical method for the design of cloaks with arbitrary boundary shapes, and apply it to faceted shapes. Subsequently, we reduce the complexity of the metamaterials needed to implement the perfect faceted cloak by introducing several approximations, whose validity is quantified by an investigation of the scattering cross section.
\end{abstract}

% Keywords required only for MST, PB, PMB, PM, JOA, JOB
%\maketitle

\section{Introduction}
In recent years, the concept of transformation optics (TO) has emerged as a powerful tool for the
control and manipulation of light~\cite{pendry2006,rahm2008}. Based on the form-invariance of
Maxwell's equations under coordinate transformation, TO is a tool in which the design of
electromagnetic materials can be performed conceptually by applying a coordinate transformation to
modify the trajectories of waves. By applying the desired coordinate transformation to Maxwell's
equations, the prescription for a medium can be obtained for which light propagates as if it was
propagating in a different coordinate system~\cite{kundtz_smith_ieee10}. One of the more compelling
concepts to emerge from the TO approach has been that of
cloaking~\cite{schurig2006,cai2007,landy2010,leonhardt2011}. The TO cloak arises from a
transformation in which a region of space is effectively shrunk to a point or singularity, where
its scattering becomes significantly reduced. The effect of the transformation is that waves appear
to be guided around the ``cloaked" region of space, rendering both the bounding TO medium and the
cloaked region invisible to an external observer. Since the first experimental realization
demonstration of a metamaterial cloak~\cite{schurig2006}, cloaking devices have been proposed for
nearly any imaginable geometry, including spheres~\cite{novitsky2009,qiu2009}, circular
cylinders~\cite{schurig2006,ruan2007,kwon2008}, cylinders of square
cross-section~\cite{rahm2008b,popa2010} and also asymmetric and irregular
shapes~\cite{han2010,nicolet2008,li2008,jiang2008,wang_zhou10}.

Though the coordinate transformations that lead to TO media are often arrived at intuitively and
can be simply described, the physical implementation of TO media are typically challenging. In
fact, while the mathematics of TO has been known for more than a century, TO has only been deemed
relevant in the context of the ongoing development of artificially structured metamaterials over
the past decade~\cite{soukoulis_wegener11}. Metamaterials are artificial media, often consisting of
arrays of metallic inclusions whose size and spacing are significantly smaller than the free space
wavelength. Under these conditions, metamaterials can be described using effective constitutive
parameter tensors. TO media are specified by spatially varying distributions of the permittivity
and permeability tensors that derive from coordinate transformations. To implement these
distributions using metamaterials, the continuously varying constitutive parameters are discretized
throughout space and an appropriate metamaterial design is chosen that achieves the desired
constitutive tensor elements at each discrete spatial point. Finding the appropriate metamaterial
element at each point, however, is a challenging task since nearly arbitrary control is required
over each of the electric and magnetic responses along six principal axes. For a given
polarization, the number of controlled responses reduces to three for a complete TO solution, which
still represents a significant challenge.

A reduction in the complexity of a TO structure can be obtained if some of the performance is
sacrificed. For example, the trajectory of a wave can be managed solely by a spatially varying,
anisotropic index of refraction~\cite{schurig2006,urzhumov2010b}. By giving up the ability to
control the wave impedance of a medium, which implies a loss of control over reflection, one gains
much more freedom in the transformation and hence the material properties. For example, the need
for magnetic (or electric) response can be eliminated if only the refractive index is to be varied.
This form of approximation has been termed the ``eikonal" approximation~\cite{urzhumov2011}.
Eikonal cloaks and other TO structures have been introduced that have more potential routes to
implementation, including the use of photonic crystals~\cite{urzhumov2010b}.

\begin{figure}
\begin{center}
       \includegraphics[width=0.4\textwidth]{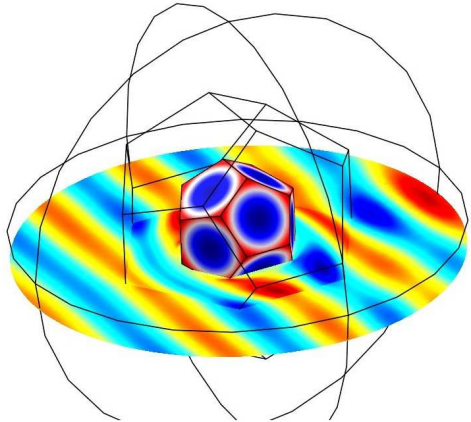}
        \caption{(Color online) Full-wave simulation: faceted dodecahedral approximation to the full-parameter spherical cloak, obtained by truncating the sphere circumscribing a regular dodecahedron.
        No correction to the material property distribution was made.}
    \label{fig:3d_cloak}
\end{center}
\end{figure}

The experimental implementation is further complicated if the cloak possesses curved contours. A
conformal cloak designed to conceal a region of arbitrary shape generally leads to a curvilinear
alignment of the principal axes of the permittivity and permeability tensors, resulting in a
gradual rotation of the local metamaterial elements throughout space. The metamaterial architecture
for such designs can be surprisingly complex in all but the most symmetric designs. Even for highly
symmetric designs, such as that of a spherical metamaterial, the fabrication can pose a significant
hurdle. Since most fabrication techniques are suited for the production of planar samples, it is of
considerable advantage to seek approximation methods that will leverage standard and commercial
lithographic patterning~\cite{soukoulis_wegener11}. For TO devices operating at
microwave~\cite{schurig2006} and THz~\cite{tao_averitt08} frequencies, cylindrical device shapes
can be obtained by curling planar sheets of metamaterials. However, as the operational frequency
increases towards the optical range and the metamaterial layer thickness decreases to micro- and
nanometer scale, this approach becomes impractical. As for TO devices with fully three-dimensional
shapes, such as spherical cloaks, curving planar-layer metamaterials is not a viable approach even
for microwave frequencies. While substantial progress is being made towards volumetric
three-dimensional metamaterial fabrication methods~\cite{soukoulis_wegener11}, the contemporary
techniques remain prohibitively difficult to use for structures that cannot be conceived as
``sandwiched" flat layers of meta-elements. For three-dimensional TO devices such as the one shown
in figure~\ref{fig:3d_cloak} faceted approximations are therefore essential.

The route to more feasible TO media considered here is the substitution of an arbitrary TO
structure, which may have curved surfaces, with a polygonal (in two dimensions) or polyhedral (in
3D) approximation. To apply the method, we first start with an exact transformation of a
two-dimensional TO device of arbitrary shape. Next, the resulting structure---which may have curved
surfaces---is approximated by polygons with an increasing number of vertices. Clearly, as the
number of vertices increases, the error in the approximation should decrease, eventually becoming
negligible. The tradeoff is to find the polygon approximation with the minimum number of vertices
that will provide acceptable performance; such a structure will be achievable by assembling a
limited number of planar sections together. The metamaterial element is identical in each planar
layer, both in terms of shape as well as orientation.

As a secondary refinement to the faceted approximation, the specific orientation of the
metamaterial elements within each planar layer can be properly taken into account by using a
transformation that maps exactly to the faceted shape. For such media, metamaterial elements whose
principal axes are coincident with the planar layer can be readily implemented, whereas other
metamaterial elements whose principal axes are not coincident with the planar layer will
necessarily complicate the fabrication steps or even preclude a proper fabrication of the TO
device.

The metamaterial invisibility cloak provides a useful structure for investigating the performance
of faceted approximations, because it can be well-characterized by a single figure-of-merit: the
(total) scattering cross-section (SCS)~\cite{kundtz_smith10,urzhumov2011}. For this reason, we
choose the cloaking transformation as the illustrative example, testing the performance of a
variety of faceted approximations.
%In the first part of the paper, we discuss two different approaches for the design of cloaks with arbitrary shapes. The proposed methods are quite general and can be adapted to irregular, convex and non-convex shapes, possibly including contours with sharp corners. In the second part, we explain the approximation scheme for TO components of arbitrary shapes for the example of a cylindrical cloak. We evaluate and quantify the different approximation steps by calculating the total SCS of the approximated cloak and compare the results to the SCS of the exact, circular-cylinder TO cloak. We discuss in detail the dependence of the SCS on the geometric and applied approximation parameters including the size of the cloak, the number of polygon surfaces and the incident angle of the wave when the approximated cloak is rotated.
%This paper is organized as follows.
In Section~\ref{sec:arbitrary_shape}, we introduce two coordinate transformation approaches for the design of arbitrary-shape cloaks.
The proposed methods are quite general and can be applied to irregular, convex and non-convex shapes, including to some extent shapes with sharp corners. We show that these methods can be used to design ideal cloaks with flat, polygonal surfaces.
In Section~\ref{sec:approx}, we introduce a polygonal approximation scheme for TO devices of arbitrary shapes, using the
cylindrical cloak of invisibility as a test bed application. We evaluate and quantify the effect of the different approximation steps by calculating the total scattering cross section (SCS) of the approximate cloaks, and compare the results to the SCS
of the exact TO cloak. We discuss in detail the dependence of the SCS on the geometric parameters and the parameters introduced by the approximations, including the size of the cloak, the number of layers, the number of sectors in the polygonal approximation,
and the angle of incidence with respect to the polygon orientation.

\section{Transformation optics of arbitrary-shape cloaks}
\label{sec:arbitrary_shape} We restrict our analysis to two-dimensional (in-plane) wave propagation
through cylindrical cloaks, possibly with non-circular cross-section. Additionally, we assume that
the incident wave is a monochromatic TE-polarized wave, that is, a wave whose electric field vector
is parallel to the cylinder axis ($z$-axis) and the wave vector is normal to it. For a cloak whose
inner and outer boundaries are circular, the exact TO solution is
well-known~\cite{schurig2006,cai2007}. For the particular case of a linear, radial transformation
of the form
\begin{equation}
r'=\frac{b(r-a)}{b-a}, \quad \phi'=\phi, \label{eq:linear_transform}
\end{equation}
the relevant tensor components of the permittivity and permeability take the form:
\begin{equation}
\epsilon_z(r) = \left(\frac{b}{b-a}\right)^2\frac{r-a}{r}, \quad \mu_r(r) = \frac{r-a}{r}, \quad
\mu_\phi(r) = \frac{r}{r-a} \label{eq:parameter_full}
\end{equation}
where $a$ and $b$ are the inner and the outer radius of the cloak, respectively.

As described in the introduction, in certain circumstances one may trade the perfectly
non-reflecting properties of ideal TO structures for a reduced fabrication burden. In the eikonal
approximation, only the refractive indexes given by $n_\phi=\sqrt{\epsilon_z\mu_r}$ and
$n_r=\sqrt{\epsilon_z\mu_\phi}$ are relevant to the wave propagation inside the
cloak~\cite{urzhumov2010}. An eikonal cloak can be derived from an exact cloak, such as the one
given by~(\ref{eq:parameter_full}), in an infinite number of ways. For example, one can fix
$\mu_\phi \equiv 1$, in which case the cloak governed by a linear transformation is described by
the following constitutive parameters~\cite{schurig2006,cai2007}:
\begin{equation}
\epsilon_z = \left(\frac{b}{b-a}\right)^2, \quad \mu_r(r) = \left(\frac{r-a}{r}\right)^2, \quad
\mu_\phi = 1. \label{eq:parameter_redu}
\end{equation}
With the reduced parameter set, the ray trajectories inside the cloak are the same as in the ideal
cloak (\ref{eq:parameter_full}), however the surface of the cloak may have a nonzero reflectance due to the impedance mismatch~\cite{schurig2006,cai2007b}. To mitigate the wave reflectance on the surface~\cite{schurig2006,cai2007b}, additional steps can be taken~\cite{cai2007b,urzhumov2010}.
%At a given frequency, the impedance mismatch can be eliminated on the surface of an eikonal cloak either by choosing a proper constant value of the relevant out-of-plane component of $\epsilon$ or $\mu$ tensors~\cite{urzhumov2010}, or by selecting a coordinate transformation that is continuous with its first derivative, in which case the refractive index, as well as wave impedance, are continuous across the outer boundary of the cloak~\cite{cai2007b}.
The advantage of using the reduced parameter set~(\ref{eq:parameter_redu}) is that only the radial
permeability~$\mu_r(r)$ varies throughout the cloak, significantly simplifying the design.

The cylindrical cloaks given by~(\ref{eq:parameter_full}) and~(\ref{eq:parameter_redu}) are
cylindrically symmetric; that is, they posses continuous rotational symmetry. The underlying
coordinate transformation is also cylindrically symmetric, which is only possible if the inner and
the outer surfaces of the cloak are both rotationally invariant. Here, we present a general
approach to cloaks whose bounding surfaces do not possess such a high symmetry, and, in fact, may
not possess any symmetry at all.

First, assume that the inner and the outer boundaries of the cloak are cylinders whose
cross-sections can be parameterized in polar coordinates as \be r=R_{1,2}(\phi),
\label{eq:angular_param} \ee where the indexes 1 and 2 refer to the inner and the outer surfaces,
respectively. For example, regular polygons with $N$ sides can be parameterized as follows:
\begin{equation}
r=\frac{r_i}{\cos(\phi \mod \Phi - \Phi/2)},
\label{eq:polygon}
\end{equation}
where $\Phi=2\pi/N$ and $r_i$ is the inradius of the polygon, i.e. the radius of the inscribed
circle of the polygon. The cloak domain (where the constitutive parameters differ from vacuum)
occupies the area described in cylindrical coordinates by $R_1(\phi)<r<R_2(\phi)$. The coordinate
transformation that compresses the cross-section of the inner surface to zero can be chosen as
follows: \be r' = \frac{R_2(\phi)}{R_2(\phi)-R_1(\phi)}\left(r-R_1(\phi)\right), \quad \phi'=\phi.
\label{eq:angular_param_transform} \ee This transformation can be rewritten in Cartesian
coordinates using $r=\sqrt{x^2+y^2}$ and $\phi=\arctan(y/x)$. Since the transformation
(\ref{eq:angular_param_transform}) depends on both $r$ and $\phi$, the $\epsilon$ and $\mu$ tensors
of the cloak are no longer aligned with the $\hat r$ and $\hat \phi$ directions of the cylindrical
coordinates. Therefore, the expressions for the tensor components are equally involved in
cylindrical and Cartesian coordinates. Since our electromagnetic solver requires material tensor
input in Cartesian coordinates, the formulas below are presented in Cartesian form.

\begin{figure}
\centering
\begin{tabular}{c}
\includegraphics[width=\columnwidth]{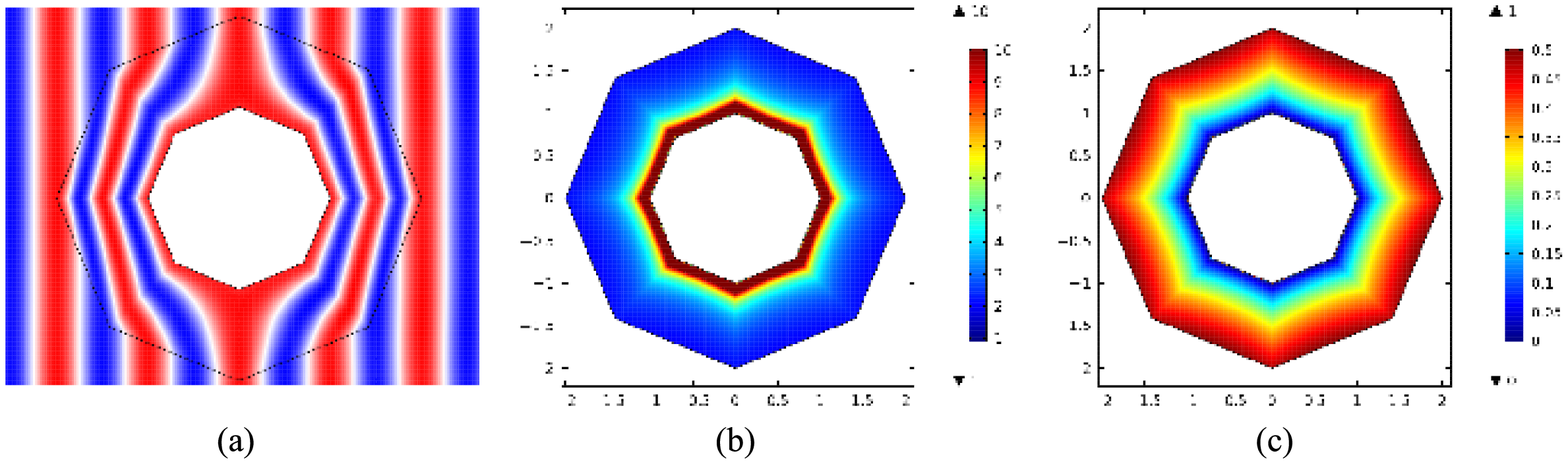}
%\includegraphics[width=0.33\columnwidth]{cloak_2d_TE_IM_octagon_plotEz}&
%\includegraphics[width=0.33\columnwidth]{cloak_2d_TE_IM_octagon_plotmu1}&
%\includegraphics[width=0.33\columnwidth]{cloak_2d_TE_IM_octagon_plotmu2}\\
%(a)&(b)&(c)\\
\end{tabular}
\caption{(Color online) The exact transformation optics octagon-shaped cloak based on the
parametrization given by (\ref{eq:polygon},\ref{eq:angular_param_transform}). (a) Field plot
($E_z)$ for a cloak illuminated by a plane TE wave.
(b,c) Distribution of the first (b) and second (c) principal values of the in-plane $\mu$ tensor in the
cloak. } \label{fig:octagon_cloak}
\end{figure}

By using $x'=r'\cos(\phi')$ and $y'=r'\sin(\phi')$, we can re-express the
transformation~(\ref{eq:angular_param_transform}) in Cartesian coordinates as $x'=x'(x,y)$ and
$y'=y'(x,y)$. In terms of the components of the Jacobian matrix of the transformation, defined as
$A_{ij}=\partial x_i'/\partial x_j$ with $i,j=1,2$ (assuming $x_1\equiv x$ and $x_2\equiv y$), the
constitutive parameters are then given by~\cite{kundtz_smith_ieee10}: \ba
\mu_{xx}=(A_{xy}^2+A_{yy}^2)/\det(A),\nonumber \\[1mm]
\mu_{xy}=\mu_{yx}=-(A_{xx}A_{xy}+A_{yx}A_{yy})/\det(A),\nonumber \\[1mm]
\mu_{yy}=(A_{xx}^2+A_{yx}^2)/\det(A),\nonumber  \\[1mm]
\epsilon_{zz}=\det(A), \label{eq:exact_TE_cloak} \ea where $\det(A)=A_{xx}A_{yy}-A_{xy}A_{yx}$. An
example simulation of a cloak whose inner and outer boundaries are shaped as regular octagons is shown in figure~\ref{fig:octagon_cloak}.

The approach described by (\ref{eq:angular_param_transform},\ref{eq:exact_TE_cloak}) is applicable
to a broad range of shapes, including non-convex shapes, with the only requirement being that the
boundaries can be parameterized in cylindrical coordinates in the form (\ref{eq:angular_param}).
For arbitrary shapes, it may be difficult to come up with an explicit analytical parametrization
such as the equation given in the polygon example (\ref{eq:polygon}). For such general shapes, we
have developed a general methodology based on the {\it wall distance} calculation, a feature
available in the COMSOL finite-element solver package~\cite{comsol}. This methodology can be
referred to as the {\it cloaking layer modeling}.

Suppose the surface of the cloak (for concreteness, the inner surface), is drawn within the
simulation software, or perhaps, generated from a CAD file. The parametrization of the surface used
internally in the software is irrelevant to our technique. In COMSOL, that inner surface can be
declared as a {\it wall}, and the distance~$D$ from any point in space to the nearest point on that
surface, along with the unit vector~$N$ pointing to it, can be found by solving the wall distance
equation~\cite{comsol}. The wall distance equation is a variation of the eikonal equation; the
equidistant surfaces thus calculated can be understood as constant-phase fronts emanating from a
fixed-phase boundary. We emphasize that this is merely an approach to parameterizing the cloak
inside its volume, and the fact that the eikonal equation is chosen to assist with that has nothing
to do with the physical approximations used. Specifically, the structure  we are reporting here is
an exact, not an eikonal, cloak.

\begin{figure}
\centering
\begin{tabular}{c}
\includegraphics[width=0.85\columnwidth]{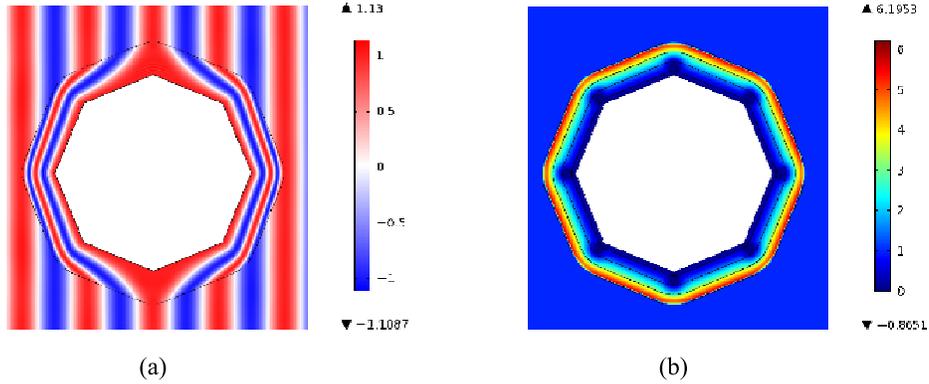}
%\includegraphics[width=0.4\columnwidth]{cloak_TE_octagon_plotEz}&
%\includegraphics[width=0.4\columnwidth]{cloak_TE_octagon_plotepszz}\\
%(a)&(b)\\
\end{tabular}
\caption{ (Color online) The octagon-shaped cloaking layer based on the parametrization given by
(\ref{eq:wall_dist_transform}) and the wall distance equation in COMSOL. (a) Field plot ($E_z)$ for
a cloak illuminated by a plane TE wave. (b) Distribution of $\epsilon_{zz}$ in the cloak. }
\label{fig:octagon_cloak_walldist}
\end{figure}

After solving the wall distance equation, the scalar field $D(x,y)$ and the vector field $\vec N
=-\grad(D)/|\grad(D)|$ are known everywhere, including the volume of the cloaking layer. Using
these fields, the cloaking transformation can be written in Cartesian coordinates as follows: \ba
x'=\frac{D}{T}\left(x+N_x(T-D)\right),\nonumber \\
y'=\frac{D}{T}\left(y+N_y(T-D)\right), \label{eq:wall_dist_transform} \ea
where $T$ is the
thickness of the cloaking layer. The inner surface of the cloak corresponds to $D=0$; in the
transformed space its size is reduced to zero. The outer surface of the cloak is an isocontour of
the wall distance, $D=T$; this surface remains the same in the transformed space ($x'=x,\,y'=y$).
Due to the nature of the wall distance equation, it has to be solved with an additional artificial
diffusion term, which, besides improving the solver convergence, provides automatic corner
smoothing as a side effect. This side effect can be seen as an advantage, if one has to design a
cloaking domain with a non-smooth inner boundary having some corners or edges, and a smooth outer
boundary; the wall distance method can provide a continuous transition between such shapes.
However, when flat faces joining at a well-defined angle are required both on the inside and
outside of the cloak, automatic corner smoothing is an unwanted effect. An example of a cloaking
layer grown on the surface of an octagon is shown in figure~\ref{fig:octagon_cloak_walldist}, along
with the field distribution of a TE-polarized wave passing through this structure.

Comparing the two TO methods presented in this section, we conclude that the method in which the
cloak domain is split into azimuthal sectors and the transformation is applied to each sector is
more suitable for shapes with corners. The piece-wise nature of such transformation preserves the
exact shapes of the corners. The second method, based on the wall distance equation, is more
appropriate for TO devices with smooth boundaries that do not contain sharp corners or edges.

\section{Polygonal approximation of arbitrary-shape cloaks}\label{sec:approx}
TO devices can often be realized by collections of metamaterial elements, each designed to provide
a local response that approximates the permittivity and permeability tensors at that point. The
first step, then, towards implementation of the TO device is the spatial discretization of the
otherwise continuous permittivity and permeability distributions. After having accomplished a
reasonable discretization, the values of the permittivity and permeability tensors can then be
realized by design of the metamaterial elements. While the individual design of the various
metamaterial elements would appear straightforward, the spatial arrangement of the elements within
the metamaterial represents an impediment to the manufacturability of the TO device.

In the following, we show that the fabrication of TO devices can be significantly simplified if the
volume of the TO device is subdivided into flat-faced subsections. In a subsequent refinement to
the faceted approximation, the electromagnetic material tensors and the corresponding metamaterial
architecture are adequately approximated and simplified within each layer. The proposed method is
general and can also be applied to three-dimensional TO objects.

\subsection{Description of the approximation methods}
\begin{figure}[t]
\begin{center}
       \includegraphics[width=0.65\columnwidth]{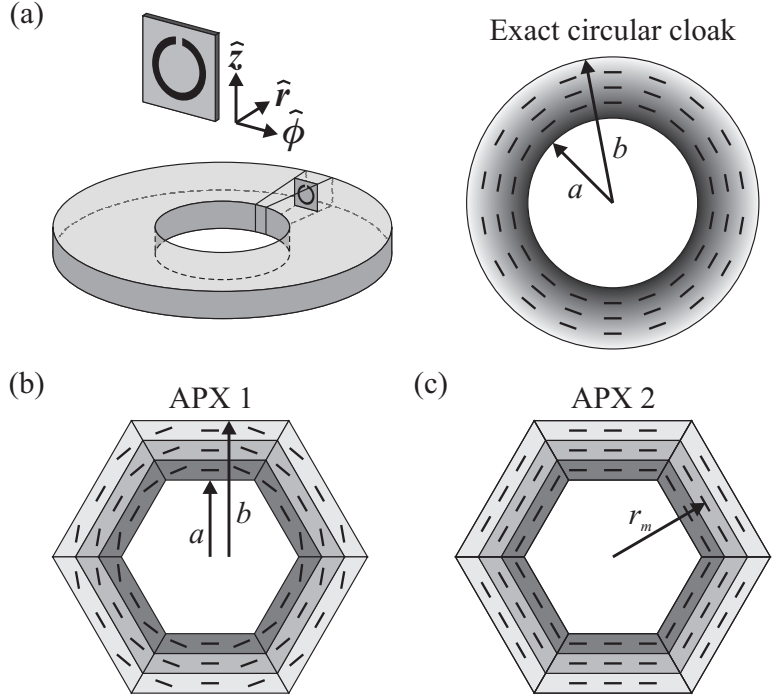}
        \caption{(a)~Exact circular cloak consisting of metamaterial elements arranged in concentric
        circles. (b) and (c)~Polygonal cloak shell approximations. In the figures, the grey shading depicts the variation of the geometric dimensions of the metamaterial elements, whereas the short lines indicate the orientation of these elements.}
    \label{fig:approx_scheme}
\end{center}
\end{figure}

Starting from either the exact cloak defined by the constitutive elements of
equation~(\ref{eq:parameter_full}), or from the eikonal cloak with the reduced constitutive
elements of equation~(\ref{eq:parameter_redu}), we investigate the impact of faceting.
Figure~\ref{fig:approx_scheme}(a) shows a schematic of the cylindrical cloak, as well as a profile
view with the metamaterial elements---here, split ring resonators (SRRs), as were used in the
microwave experiments---laid out in a circumferential pattern. The metamaterial elements are the
first level approximation to the continuous transformations. If the metamaterial element can
control all three of the relevant constitutive tensor elements at a given point, then a discrete
approximation to the full transformation is achieved. If the metamaterial element is designed to
control only two of the constitutive tensor elements, then the structure represents a discrete
approximation to the eikonal cloak.

In this approach, the cylindrical cloak consists of a certain number of concentric rings, each with
different radius $r$. The radius is measured at the radial midpoint of the concentric ring. Each
concentric ring is composed of metamaterial elements with identical values of $\mu_r$, $\mu_\phi$
and $\epsilon_z$. The material parameters can be readily obtained from numerical full-wave
simulations of a single unit cell in each concentric ring. Although there is no repeated unit cell
in the final TO structure, the retrieved effective medium properties for a metamaterial element
achieved by assuming periodicity of the element have been shown to be a fairly accurate
description, and applicable to gradient and TO media.

Figure~\ref{fig:approx_scheme}(b) and \ref{fig:approx_scheme}(c) show two different faceted
approximations of the circular cylinder cloak which we refer to as the approximations APX~1 and
APX~2, respectively. In both APX~1 and APX~2, the circular cloak is approximated by trapezoid
sectors forming a polygon. Consequently, the cloak can be realized by metamaterial elements that
are arranged into flat layers that run parallel to the exterior boundary of each sector. By
stacking these different layers, the polygonal approximation to a cylindrical cloak is established.
The flat metamaterial layers are significantly easier to fabricate using standard lithographic
techniques than those with curved boundaries.

We next describe how the $\epsilon$ and $\mu$ tensors are calculated for the metamaterial elements
in each layer in each approximation. For this purpose, we assume that the variation of the
cylindrical radius $r$ (which represents the distance from the center) can be neglected within one
metamaterial layer.

In APX~1, the continuous tensor fields $\epsilon(r)=\epsilon_z(r)\hat z \hat z$ and
$\mu(r,\phi)=\mu_r(r)\hat r\hat r + \mu_\phi(r)\hat\phi\hat\phi$ from the exact circular cylinder
cloak are approximated as
\begin{eqnarray}
\epsilon_{\mbox{\tiny{APX1}}}(r) &= \epsilon_z(r_m)\hat z \hat z, \nonumber \\
\mu_{\mbox{\tiny{APX1}}}(r,\phi) &= \mu_r(r_m)\hat r\hat r  + \mu_\phi(r_m)\hat\phi\hat\phi.
\label{eq:APX1_definition}
\end{eqnarray}
Here, $r_m$ is the radius evaluated at the mid-point of the $m$-th metamaterial layer in each
sector (see figure~\ref{fig:approx_scheme}); for example, if the sector spans between angles
$\phi_1$ and $\phi_2$, $r_m$ is evaluated at the mid-angle $\phi_m=(\phi_1+\phi_2)/2$. In APX 1,
the permittivity is thus constant in each layer and each sector, and the permeability tensor has
constant principal values but variable principal axes.

In APX~2, we additionally neglect the rotation of the principal axes of the metamaterial elements
within each sector. The material tensors are thus given by
\begin{eqnarray}
\epsilon_{\mbox{\tiny{APX2}}}(r) & =\epsilon_z(r_m)\hat z \hat z,  \nonumber \\
\mu_{\mbox{\tiny{APX2}}}(r,\phi) & = \mu_r(r_m) \hat r_m \hat r_m + \mu_\phi(r_m)
\hat\phi_m\hat\phi_m. \label{eq:APX2_definition}
\end{eqnarray}
Here, $\hat r_m$ and $\hat\phi_m$ are the unit vectors of the cylindrical coordinate system
evaluated at the mid-angle of each sector of the polygon. In APX~2, both $\epsilon$ and $\mu$
tensors are entirely constant in each layer within each sector; the orientation of the principal
axes of $\mu$ experiences a jump between sectors (see figure~\ref{fig:approx_scheme}(c)).

\subsection{Simulation results and discussion}
\begin{figure}
\begin{center}
       \includegraphics[width=\textwidth]{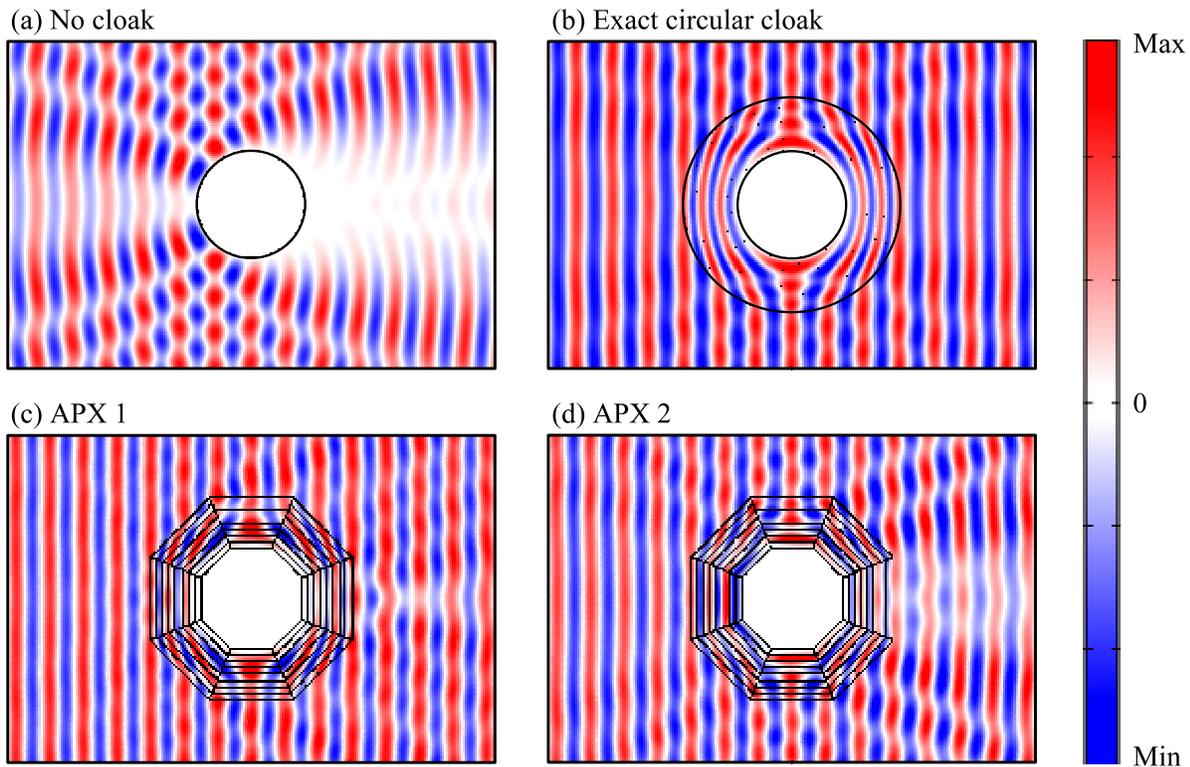}
        \caption{(Color online) Simulation results for the $z$-component of the electric field of a TE-wave incident from the left. (a)~The bare conducting object without a cloak. (b)~The exact circular-cylindrical cloak.
        (c)~Approximations APX~1 and (d)~APX~2 simulated with the full parameter set.}
    \label{fig:felder}
\end{center}
\end{figure}

In order to quantify the validity of the different approximations APX~1 and APX~2, we perform
numerical calculations and determine the scattering cross-section (SCS) for an exact cylindrical
cloak as well as the approximate realizations APX~1 and APX~2. In the model, the mesh size is
$\lambda/15$ in regions outside the cloak and $1/5$ of the layer thickness (but at most
$\lambda/15$) inside the cloak shell. The cloak core is filled with a quasi-perfect electrical
conductor with a conductivity of $\unit[10^{16}]{S/m}$. To verify the appropriateness of the
numerical model, we compute the field distribution for an incident plane wave for the uncloaked
core, the exact cloak and for the two approximations APX~1 and APX~2, respectively (see
figure~\ref{fig:felder}(a) to (d)). The plots confirm that the two approximations reduce both the
backscattering as well as the forward scattering, leading to a field pattern that is qualitatively
comparable to that of the exact cloak.

For a quantitative description of the proposed approximations, we calculate the SCS of the
different cloaks for an incident plane wave~\cite{urzhumov2011}. According to the optical theorem
in two dimensions~\cite{boya_murray94}, the total cross-section (which includes absorption and
scattering) can be determined from the forward scattering amplitude $f(0)$:
\begin{equation}
\sigma_{tot} \equiv  \sigma_{sc} + \sigma_{abs} = -2\sqrt{\lambda} \,\mbox{Re}\{\sqrt{i} f(0)\},
\label{eq:optical_theorem}
\end{equation}
where $\lambda$ is the wavelength in the medium surrounding the scatterer---in our case, the free
space wavelength. The scattering amplitude depends only on the scattering angle and is defined by
the asymptotic formula \be E(r,\phi) \sim E_0\left(e^{ikr} + \frac{f(\phi)}{\sqrt{r}}e^{ikr}
\right)$, \quad $r\to\infty, \ee in which $r$ is the distance from the scatterer, $E_0$ is the
amplitude of the incident plane wave, and $k=2\pi/\lambda$ is the wave number. Thus, in two
dimensions $f(\phi)$ has the units of (length)$^{1/2}$, and $\sigma_{tot}$ has the units of length.
For a lossless scatterer, $\sigma_{abs}=0$ and thus the optical theorem gives the SCS
($\sigma_{sc}$) directly without additional calculations. Our simulation software permits the
calculation of the far-field amplitude in the forward scattering direction; that quantity is
linearly proportional to the forward-scattering amplitude in equation~(\ref{eq:optical_theorem}),
which permits us to perform an inexpensive calculation of the total SCS at each frequency.

In the following, we present the SCS for the two approximations APX~1 and APX~2 as a function of
the size of the cloaked area, the cloak shell thickness, the number of polygon sides and layers and
the angle of wave incidence.

To evaluate the limitations of the proposed models, we first calculate the SCS as a function of the
size of the cloak. For this purpose, we vary the radius of the cloaked area from $a=0.5\lambda$ to
$a=3.5\lambda$ while the outer radius of the cloak is proportionally increased according to $b=2a$.
The corresponding material parameters of the full parameter set (see (\ref{eq:parameter_full}))
are:
\begin{equation}
0\le \epsilon_z\le 2, \quad 0\le\mu_r\le0.5, \quad 2\le \mu_\phi \le \infty.
\end{equation}
The diverging behavior of $\mu_\phi$ occurs only near the inner boundary of the cloak where the
electromagnetic fields are negligibly small. Therefore, in an explicit metamaterial implementation,
the limited nature of the effective permeability produces only a small error in this region. For
the reduced parameter set of (\ref{eq:parameter_redu}), however, all material parameters remain
finite:
\begin{equation}
\epsilon_z=4, \quad 0\le\mu_r\le0.25, \quad \mu_\phi = 1.
\end{equation}
Note that the parameter ranges of both the full and the reduced parameter set are independent of
the size of the cloaked area which means that all cloak sizes can be realized by the same set of
metamaterial elements.

\begin{figure}
\begin{center}
       \includegraphics[width=0.55\columnwidth]{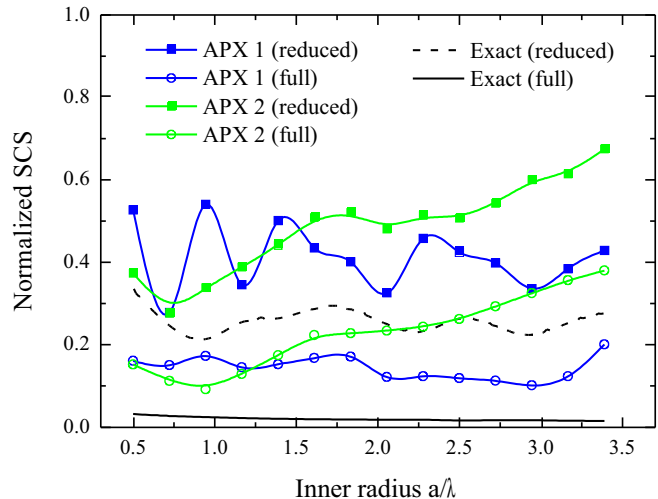}
        \caption{(Color online) Scattering cross section (SCS) normalized to the SCS of the bare conducting core for the proposed approximations APX~1 and APX~2 in
        dependence on the core radius~$a$ for a constant cloak shell ratio~$b=2a$ and a $10$-sided polygon. The black lines show the corresponding SCS of the exact circular-cylindrical cloak.}
    \label{fig:size}
\end{center}
\end{figure}

In figure~\ref{fig:size}, we show the resulting SCS, normalized to the SCS of the bare conducting
core for the two approximations and for both the full and the reduced parameter set. To achieve
sufficient cloaking even for a large cloak, we assumed a 10-sided polygon for all models.
Furthermore, the number of layers in each polygon section was step-wise increased with increasing
size of the cloak (starting with 5 layers) to keep the layer thickness smaller than $\lambda/10$.
For comparison, we have also plotted in figure~\ref{fig:size}, the SCS of the exact
circular-cylindrical cloak for the full and the reduced parameter set, respectively (solid and
dashed black lines).

As expected, the models that are based on the full parameter range (indicated by small circles)
generally provide a better cloaking ability than the models with reduced parameter set (small
squares) with a reduction of the SCS of about a factor of two. For the considered cloak core sizes,
the SCS of model APX~1 is nearly independent of the size and reaches values around SCS $=0.15$ if
the model is initialized with the full parameter set (blue line with circles) and SCS $=0.41$ for a
cloak based on the reduced set (blue line with squares). In contrast to this stable behavior, the
SCS of the model APX~2 slightly increases with increasing core size and approaches values of about
$0.45$ for the full parameter set and $0.70$ for the reduced set as the core size approaches an
inner radius of $a=3.5\lambda$. The increasing SCS for the model APX~2 is a direct consequence of
the stronger approximations made in this model (as shown later, this can be compensated by
increasing the number of polygon edges).

\begin{figure}
\begin{center}
       \includegraphics[width=0.55\columnwidth]{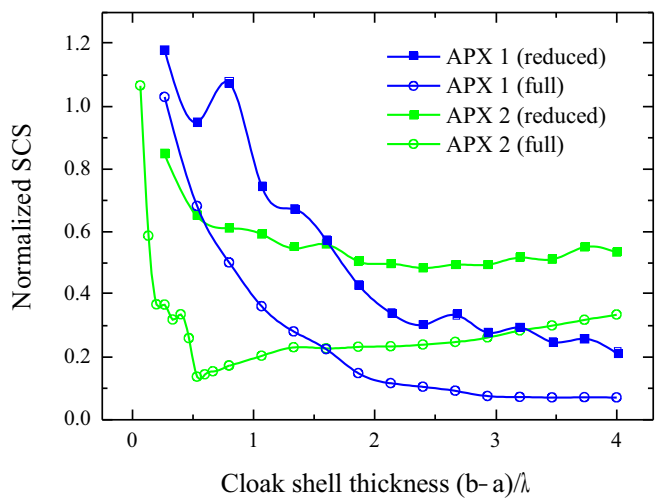}
        \caption{(Color online) Scattering cross section (SCS) normalized to the SCS of the bare conducting core for the proposed approximations APX~1 and APX~2 in
        dependence on the cloak
shell thickness for a constant core radius $a=2\lambda$ and a $10$-sided polygon.}
    \label{fig:shell}
\end{center}
\end{figure}

Once the area to be cloaked is specified, the cloak can still be optimized by varying the size of
the cloak shell. Therefore, we determined the cloaking performance of the different approximations
with respect to the thickness of the cloak shell $(b-a)$ for a constant inner radius of
$a=2\lambda$ and for a 10-sided polygon. As before, the number of layers in each polygon section
was stepwise increased with growing outer radius~$b$ such that the thickness of each layer was
$\lambda/10$ at most. The resulting SCS for the two approximations in dependence on the normalized
shell thickness $(b-a)/\lambda$ is plotted in figure~\ref{fig:shell}.

The total dependence of the SCS on the cloak shell thickness is a combination of several effects.
First, if the cloak shell is very thin, the distribution of the material parameters is compressed
into a small region of space. This leads to a large jump of the material parameters at the
interface between adjacent layers (in which the parameters are constant) and thus to an increased
scattering of light. Consequently, the SCS for all models gets very large as the cloak shell
thickness tends to zero. Second, for the models that are based on the reduced parameter set, the
wave impedance of the outer cloak boundary $z = \sqrt{\mu_\varphi/\epsilon_z} = 1-a/b$ (see
(\ref{eq:parameter_redu})) approaches that of free space as the outer boundary $b$ increases ($a$
is fixed). This leads to a decreasing scattering. On the other hand, if the shell thickness
increases, the role of spatial imperfections, i.e.~the deviations from the exact material
parameters at a given point, gain significance and lead to an enhanced scattering. This is
especially the case for the model APX~2 which includes stronger approximations than model APX~1.

\begin{figure}[t]
\begin{center}
       \includegraphics[width=\textwidth]{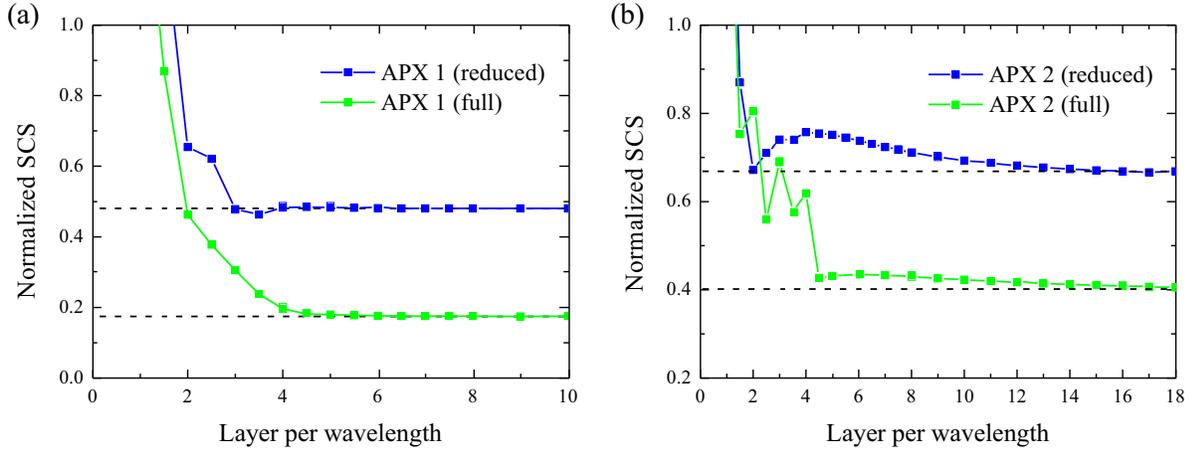}
        \caption{(Color online) Normalized scattering cross section (SCS) in dependence on the package density of the metamaterial layers simulated with the full parameter set (red dotted line) and the reduced (black dotted line) parameter set.}
    \label{fig:conv}
\end{center}
\end{figure}

For the considered realizations, the net effect for the model APX~1 is a general decrease of the
SCS for increasing shell thicknesses (see blue lines in figure~\ref{fig:shell}). Obviously, within
the considered thickness range, the improved impedance matching for larger cloaks exceeds the
adverse effect of the increasing spatial imperfections. The corresponding SCS approaches a minimal
value of $0.21$ for the reduced parameter set and $0.07$ for the full parameter set, respectively.
In contrast, for the model APX~2 the impact of spatial imperfections becomes more crucial. For the
full parameter set (green line with small circles in figure~\ref{fig:shell}), the imperfections
start to overcompensate the improved impedance matching at a shell thickness of $0.5\lambda$
leading to a local scattering minimum where the SCS decreases to $0.14$. Yet, the corresponding
curves of the reduced parameter set show a more stable behaviour with an almost
thickness-independent SCS around $0.6$ (see green line with small squares in
figure~\ref{fig:shell}). Here, the two counter effects of the increasing imperfections and the
improving impedance matching are almost balanced.

%\subsection{Variation of the number of layers}
In order to estimate the fabrication effort of the layered cloaks, we next investigate the number
of layers that are necessary to provide a reasonable cloaking effect. In figure~\ref{fig:conv}(a)
and (b) we show the normalized SCS as a function of the number of layers for model APX~1 and APX~2
for both the reduced and the full parameter set, respectively. The assumed geometry of the cloak
was an 8-sided polygon with an inner radius of $a=2\lambda$ and an outer radius of $b=4\lambda$.

The curves show a fast convergence towards a constant SCS which corresponds to a quasi-continuous
distribution of the material parameters within the cloak. Obviously, a sufficient convergence is
already achieved for a package density of 4~layers per wavelength for model APX~1
(figure~\ref{fig:conv}(a)) and for 6~layers per wavelength for model APX~2
(figure~\ref{fig:conv}(b)). Below this limit, the major contribution to the SCS arises from
internal reflections between adjacent layers. In this regard, the full and the reduced parameter
set lead to a similar SCS of the cloak. In other words, the higher effort of using metamaterial
elements that provide a full parameter set is only reasonable if the thickness of the layers is
smaller than~$\lambda/4$ for model APX~1 and smaller than~$\lambda/8$ for model APX~2,
respectively. But this weak condition is usually satisfied in common metamaterial implementations.

%\subsection{Variation of the number of polygon edges}
Next, we determine the number of necessary polygon sides in dependence on the area to be cloaked.
For short, we exemplarily show this dependence only for the reduced parameter set of model APX~2
since this is the most interesting configuration for a practical realization. As before, the outer
radius~$b$ was proportionally increased with increasing inner radius~$a$ according to $b=2a$ to
ensure that the same material parameter range is covered as in the preceding discussion, and we
assumed a layer thickness of about~$\lambda/10$.

\begin{figure}
\begin{center}
       \includegraphics[width=0.6\columnwidth]{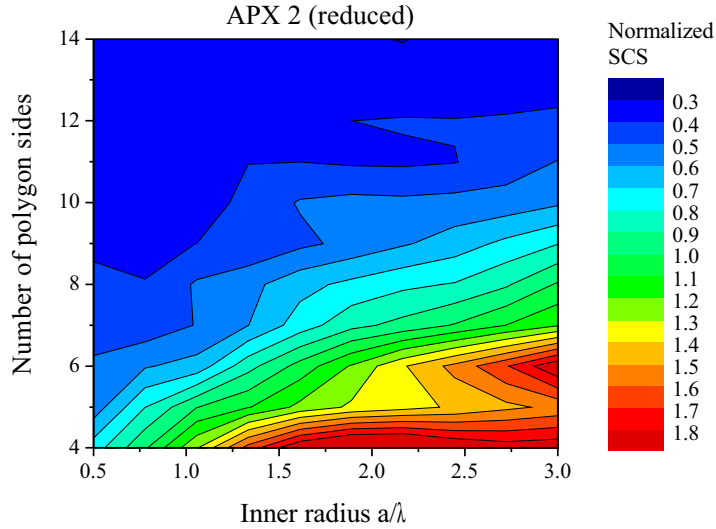}
        \caption{(Color online) Normalized scattering cross section (SCS) in dependence on the inner radius~$a$ of the cloak and the number of polygon sides for the approximation APX~2. The simulations are based on the reduced parameter set and refer to a cloak with an outer radius of $b=2a$.}
    \label{fig:size_sides}
\end{center}
\end{figure}

The resulting normalized SCS in dependence on the inner radius~$a$ and the number of polygon
sides~$N$ is shown as a 2D plot in figure~\ref{fig:size_sides}. As expected, the number of
necessary polygon sides increases for an increasing cloak size. For example, if we intend to reduce
the SCS of the cloak below 50\,\% of the SCS of the uncloaked core, the minimum number of polygon
sides is $N=7$ for an inner radius of $a=0.5\lambda$ and $N=11$ for $a=3\lambda$. For more than 12
polygon sides, there is no significant further reduction for all analyzed sizes.

\begin{figure}[t]
\begin{center}
       \includegraphics[width=\textwidth]{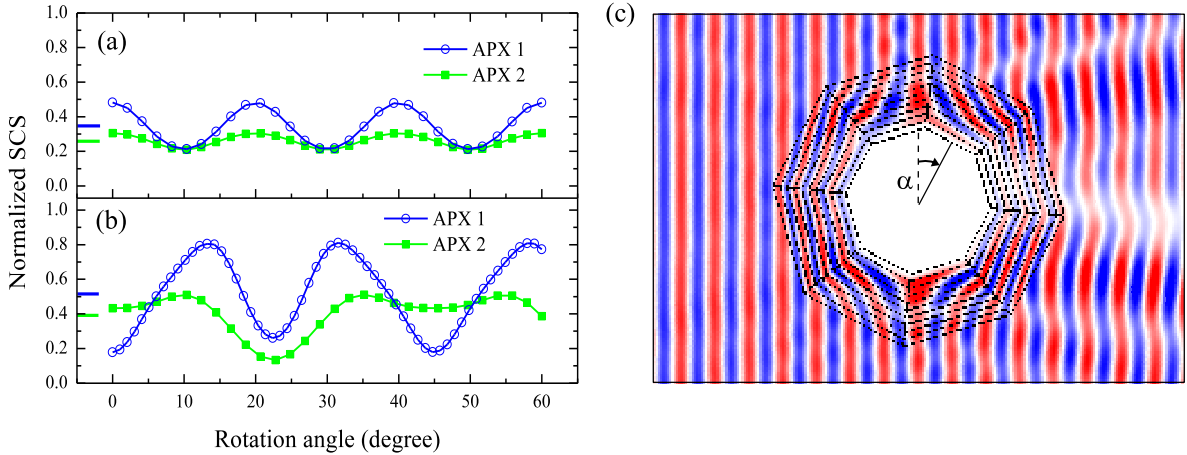}
        \caption{(Color online) (a)~Normalized scattering cross section (SCS) in dependence on the angle of
        incidence
        for a 9-sided polygonal cloak calculated for the full parameter set. The angle of $0^\circ$ corresponds to the case where the wave incidence is normally to a polygon side. (b)~Same as (a), but for an 8-sided polygon. The short lines at the $y$-axis
        indicate the annularly averaged SCS. (c)~$z$-component of the scattered electric field, exemplarily shown for an incident angle of
$17^\circ$ for the 8-sided version of APX~2.}
    \label{fig:angle}
\end{center}
\end{figure}

%\subsection{Variation of the angle of incidence}
As a final aspect of the proposed polygonal approximations, we analyze the influence of the
orientation of the cloak with respect to the incident wave. Since the polygonal shape breaks the
rotational symmetry of the original cylindrical cloak, the used approximations are expected to be
sensitive to the angle of incidence. This angle dependence is exemplarily shown in
figure~\ref{fig:angle} where we have plotted the normalized SCS of the models APX~1 and APX~2 as a
function of the angle of incidence. The calculations are based on the full parameter set and we
considered an 8-sided polygon (figure~\ref{fig:angle}(b)), as well a 9-sided polygon
(figure~\ref{fig:angle}(a)). In the simulations, we assumed an inner and outer radius of
$a=2\lambda$ and $b=4\lambda$, respectively, and 10~layers in each polygon section.

The resulting curves plotted in figure~\ref{fig:angle}(b) show a periodic modulation which mirrors
the expected 45$^\circ$ annular period of the 8-sided polygon. In contrast, the curves calculated
for the 9-sided polygon (see figure~\ref{fig:angle}(a)) display a period of only 20$^\circ$
i.e.~only the half of the expected 40$^\circ$ period of an 9-sided polygon. The reason for the
higher symmetry is the odd number of polygon sides as can be explained as follows: there are two
symmetric orientations of a polygonal cloak with respect to the incident wave. The first occurs if
a polygon side is normal to the incident wave (case~1) and the second occurs if a vertex of the
polygon points towards the source (case~2). If the polygon has an even number of sides the two
cases possess different SCS because the front and the back of the cloak are either two parallel
sides (case~1) or two opposing vertices (case~2). However, if the number of polygon sides is odd, a
side is always opposite to a vertex and thus the two cases have a comparable geometry. This leads
to similar values of the SCS and thus to a halving of the annular period.

Furthermore, the curves in figure~\ref{fig:angle} show that model APX~2 generally provides a more
stable SCS under cloak rotation than model APX~1, and also the annularly averaged SCS of APX~2 is
smaller than that of APX~1. For example, for an 8-sides polygon (see figure~\ref{fig:angle}(b)) the
mean value and standard deviation of the averaged SCS are $0.51\pm 0.21$ for model APX~1 but only
$0.39\pm 0.12$ for model APX~2.

In summary, both polygonal approximations of the exact cylindric cloak show a comparable and
reasonable cloaking performance. For a given size of the area to be cloaked, the reduction of the
SCS can be optimized by varying the size of the cloak shell and the number of polygon edges. For
the analyzed sizes ($ a \le 3.5\lambda $) and realistic geometric dimensions, we achieved a
reduction of the SCS on the order of 10\,\% of the uncloaked core. In general, model APX~1 provides
a slightly better cloaking figure-of-merit, since the SCS shows a faster decrease with increasing
cloak thickness; on the other hand, APX~2 is less sensitive to the angle of incidence. For both
models, the role of inter-layer reflections is negligible as long as the layers of metamaterial
elements are thinner than $\lambda/4$ (APX~1) or $\lambda/8$ (APX~2), respectively, which is also a
realistic assumption for a practical implementation. From the ease-of-fabrication point of view,
model APX~2 has the advantage that the metamaterial elements are all identically aligned within one
layer. Such uniform, flat layers can be manufactured even with the micro- and nanoscale dimensions
by standard fabrication methods.
%This enables the construction of
%polygonal cloaks even in the high frequency range.

\section{Conclusion}
In this paper, we have proposed an approximation method for the design of transformation-optical
(TO) components of arbitrary shape. Starting from the exact transformation, a polygonal
approximation of the curved shape of the TO device is introduced. In consecutive approximations the
metamaterial elements were aligned parallel to the flat boundaries of the polygons. As a major
advantage, the polygonal approximation significantly mitigates the fabrication constraints since
the resulting TO components can be implemented by flat metamaterial layers. This can be readily
achieved by standard lithographic and stacking techniques. We have validated the approach for the
example of a cylindrical cloak and quantified the accuracy of the approximations by comparing the
scattering cross-section of the cloak with the exact TO cloak of a circular shape.

Furthermore, we have presented an alternative method for the design of electromagnetic cloaks with
virtually arbitrary shapes. The approach is based on the \emph{wall distance} calculation and is
applicable to a large class of cloak geometries including convex shapes and even some non-convex
shapes. The new technique has been numerically verified on a specific example of an octagon-shaped
cloak.

\section*{Acknowledgments}
This work was partially supported through a Multidisciplinary University Research Initiative,
sponsored by the U.S.~Army Research Office (Contract No. W911NF-09-1-0539). The authors are
thankful to Christian Wollblad (COMSOL AB) for assistance with the wall distance equation.

\section*{References}
%\bibliographystyle{iopart-num}
%\bibliography{literatur}
\providecommand{\newblock}{}

\end{document}